\documentstyle[aps,graphicx]{revtex}

\begin{document}
\draft
\title{Sensitivity Study of Extra Dimensions at TeV $e^+ e^-$ colliders
\footnote{
To be published in the proceedings of the theory meeting on physics at 
linear colliders, 15--17 March 2001, KEK, Japan.
}}
\author{Kingman Cheung}
\address{National Center for Theoretical Science, 
National Tsing Hua University, Hsinchu, Taiwan R.O.C.}
\maketitle

\begin{abstract}
We study the sensitivity reach of $0.5-2$  TeV $e^+ e^-$ linear colliders 
in the context of the ADD model, in which gravity becomes strong at TeV scale.
We include the real emission channel $e^+ e^- \to \gamma G$, as well as
the virtual-graviton-exchange channels $e^+ e^- \to \gamma \gamma, q\bar q,
\mu^+\mu^-, \tau^+ \tau^-, e^+ e^-$. 
Assuming no excess of events over the standard model predictions, we 
obtain the lower 95\% C.L. limits on the effective Planck
scale.  These limits are better than those obtained in the Run II's of
the Tevatron and comparable to those of the LHC.
\end{abstract}

\section{Introduction}

Recent advances in string theory have stimulated many activities in
 particle phenomenology.
The previously unreachable Planck, string, and grand unification 
scales ($M_{\rm Pl}$, $M_{\rm st}$, and $M_{\rm GUT}$, respectively) can 
be brought down to a TeV range.
In this case, one expects low energy phenomenology
testable at current and future collider experiments.

An attractive realization of the above idea was proposed by
Arkani-Hamed, Dimopoulos, and Dvali (ADD) \cite{ADD}.  
In their model, the Standard
Model (SM) particles live on a D3-brane, which is a topological object in 
higher dimensional space, where open strings end.
On the other hand, gravity is allowed to propagate in the extra
dimensions.  In order to bring the Planck scale ($10^{19}$ GeV) to TeV range, 
the size of these compactified dimensions is made very large compared to 
$(M_{\rm weak})^{-1}$.  The relation among the Planck scale $M_{\rm Pl}$, 
size $R$ of the extra dimensions, and the effective Planck scale $M_S$ is:
\[
M^2_{\rm Pl} \sim M_S^{n+2} \; R^n \;,
\]
where $n$ is the number of extra (compactified) dimensions.
Assuming that the effective Planck scale
$M_S$ is in the TeV range,  it gives a very large $R$ of
the size of a solar system for $n=1$, which is obviously ruled out by
experiments. However, for all $n\ge 2$ the expected $R$ is less than 
$1$~mm, and therefore do not contradict existing gravitational experiments.

A graviton in the extra dimensions is equivalent, in
the 4D-point of view, to a tower of infinite number of Kaluza-Klein
(KK) states with masses $M_k = 2\pi k/R\;\; (k=0,1,2,...,\infty)$.
The coupling of the SM particles residing on the brane to each of 
these KK states is of order $1/M_{\rm Pl}$.
The overall coupling is, however, obtained by summing over all the KK
states, and so scales as $1/M_S$. Since the $M_S$ is in the TeV
range, the gravitational interaction is as strong as electroweak
interactions, and thus can give rise to many testable consequences in
both accelerator and non-accelerator experiments.
In general, present collider 
experiments can constrain the effective Planck scale at around 1 TeV.
For a mini-review of the collider phenomenology please see Ref. \cite{pascos}.

In this study, we estimate the sensitivity reach of the $e^+ e^-$ linear
colliders (LC) of 0.5 -- 2 TeV center-of-mass energies in the context of the 
ADD model.  We include the following processes in our consideration
\begin{eqnarray}
e^+ e^- & \to & \gamma G \nonumber \\
e^+ e^- & \to &  Z G \nonumber\\
e^+ e^- & \to & \gamma \gamma \nonumber\\
e^+ e^- & \to &  q \bar q \nonumber\\
e^+ e^- & \to & \mu^+ \mu^-\;, \;\; \tau^+ \tau^-  \nonumber\\
e^+ e^- & \to &  e^+ e^- \nonumber
\end{eqnarray}
where $G$ denotes the graviton.  The first two processes involve real emission
of the graviton, which escapes into the extra dimensions and thus gives
rise to missing-momentum events.  The other processes involve virtual-graviton
 exchanges in addition to the usual electroweak exchanges.  The 
interference between the SM amplitude and the graviton-exchange amplitude
affects the total cross section and various distributions.
If experimentally these 
distributions can be measured to a high precision, strong constraints can be
placed on the new interactions of the graviton.  In this study, we
assume that the SM is correct and experiments give cross sections and 
distributions consistent with the SM, then we use the data to constrain 
the effective Planck scale.

\section{Procedures}

\subsection{Experimental Acceptance}

We use typical kinematic and geometrical acceptance cuts for the detector in
the LC:
\begin{equation}
|\cos\theta_i| <0.9\;,\qquad  E_i > 20\; {\rm GeV} \;,
\end{equation}
where $i = e,\mu,\tau,q$, or $\gamma$.  
The center-of-mass energies are 0.5,1, 1.5, and 2 TeV and 
the integrated luminosities used in
our study are 10 to 200 fb$^{-1}$.
We employ an overall efficiency of 0.8 for all the events.

\subsection{Monte Carlo data generation}

In order to estimate the sensitivity of collider experiments to the
ADD model, we need to generate some ``realistic'' data
sets.  We use the SM cross
section to generate a smooth single or double differential
distribution.  We divide the distribution into bins:
in each bin $i$ the
expected number of events $S^{\rm SM}_{i}$ is obtained by
multiplying the cross section in this bin by the known integrated
luminosity and efficiency. We further proceed with a Monte Carlo (MC)
experiment. For each bin $i$ we generate a
random number of events $n_{ij}$ using Poisson statistics with the mean 
$S^{\rm SM}_{i}$.
We use the MC data sets generated in
this way to obtain the best fit and the limits on the gravity scale
$M_S$ for the distributions.

As a result of the fit we obtain the 95\% C.L. upper limit on $\eta$.  We
then repeat the above procedures
many times. The limits obtained in these repeating experiments are
histogrammed.  The sensitivity to the scale $M_S$ is defined as the
median of this histogram, i.e. the point on the sensitivity curve
which 50\% of the future experiments will exceed. 

\subsection{Fitting procedure}

We extract the lower limit on the gravity scale $M_S$ by fitting each of the
random MC experiments with a sum of the SM background and Kaluza-Klein
graviton contribution. We employ the maximum likelihood method with the 
fitting parameter $\eta={\cal F}/M_S^4$.

We, therefore, generate three templates that describe the cross section
in the case of large extra dimensions. The first one describes the SM
cross section. The other two describe terms
proportional to $\eta$ (interference term) and to $\eta^2$
(Kaluza-Klein term), respectively.
We then parameterize production cross
section in each bin as 
\begin{equation}
	\sigma = \sigma_{\rm SM} + \sigma_4 \eta + \sigma_8 \eta^2,
\label{eq:template}
\end{equation}
where $\sigma_{\rm SM}$, $\sigma_4$, and $\sigma_8$ are the three
templates described above.
We combine channels by adding their likelihoods.

\section{Various Processes and Results}

\subsection{Real Emission: $e^+ e^- \to \gamma G$}

This is one of the earliest processes in the discussions of collider
signatures for the ADD model \cite{eegG-ear,eegG}.  The graviton escapes to the
extra dimensions, which gives rise to a missing-energy signature.  The
differential cross section versus the cosine of the scattering angle 
is given by
\begin{eqnarray}
\frac{d\sigma}{d \cos\theta} &=& \frac{\pi \alpha G_N}{ 4 \left( 1- 
\frac{m^2}{s} \right )} \; \biggr[ (1+\cos^2\theta) \left( 1 +
(\frac{m^2}{s})^4 \right ) \nonumber \\
&&+ \left( \frac{ 1-3\cos^2\theta +4\cos^4\theta}{1-\cos^2\theta} \right )
\, \frac{m^2}{s} \, \left( 1+ (\frac{m^2}{s})^2 \right ) +
6 \cos^2\theta (\frac{m^2}{s})^2 \biggr ] \;,
\end{eqnarray}
where $s$ is the square of the center-of-mass energy and $m$ is the mass of the
graviton.  Since the spectrum of graviton states behaves like a continuous one
at the energy scale that we are considering, the mass $m$ is to be integrated
from 0 upto the center-of-mass energy of the $e^+ e^-$ collider.  The relation
of the energy of the photon to the mass of the graviton is 
\[
E_\gamma = \frac{s-m^2}{2 \sqrt{s}} \;,
\]
by which we can obtain the double differential cross section
$d^2\sigma/ d E_\gamma d\cos\theta_\gamma$. 

The irreducible background comes from $e^+ e^- \to \gamma \nu \bar \nu$.
The main difference between the signal and background is the $Z$
peak of the background in the recoil-mass distribution.  We used the 
following cuts to reduce the background \cite{eegG}:
\begin{equation}
\label{cuts}
M_{\rm recoil} > 200 \; {\rm GeV}\;, \qquad E_\gamma > 20\;{\rm GeV}\;,
\qquad |\cos\theta_\gamma|<0.9 \;.
\end{equation}

The process $e^+ e^- \to Z G$
 is similar to the previous one with the $\gamma$ replaced by $Z$.
The $Z$ boson decays into a lepton pair or a quark pair.  The irreducible
background comes from $e^+ e^- \to Z \nu \bar \nu$.  The cuts to reduce the
background are similar to those in Eq. (\ref{cuts}).  The formula for this
process can be found in Ref. \cite{eegG}.  Since the
limits that can be obtained from this channel is inferior to the $e^+ e^- \to
\gamma G$ channel, we shall ignore this channel in obtaining the combined 
limits.
In the estimation of the sensitivity reach, we used the two-dimensional 
distribution of $d^2\sigma/d E_\gamma d\cos\theta_\gamma$.  The reach by this
channel is shown in Table \ref{table1}.

\begin{table}[th]
\caption{\small \label{table1}
95\% C.L. limit on $M_S$ (TeV) using the real graviton-emission channel
$e^+ e^- \to \gamma G$}
\medskip
\begin{tabular}{c||cc|cc}
& \multicolumn{2}{c|}{$\sqrt{s}=0.5$ TeV} & 
\multicolumn{2}{c}{$\sqrt{s}=1$ TeV}  \\
 $n$ & ${\cal L}=10$ fb$^{-1}$ & 50 fb$^{-1}$ & 
 ${\cal L}=10$ fb$^{-1}$ & 50 fb$^{-1}$ \\
\hline
\hline
2   & 4.7 & 5.7 & 7.2 & 8.8\\
3   & 2.8 & 3.4 & 4.6 & 5.4 \\
4   & 2.1 & 2.4 & 3.5 & 4.0\\
5   & 1.7 & 1.9 & 2.9 & 3.3\\
6   & 1.4 & 1.6 & 2.5 & 2.8 \\
\end{tabular}
\end{table}

\subsection{Virtual Graviton Exchange}
\subsubsection{$e^+ e^- \to \gamma\gamma$}

The differential cross section versus $z=|\cos\theta|$ 
($\theta$ is the polar angle of the outgoing photon) is given by \cite{eegg}
\begin{equation}
\label{ee}
\frac{d\sigma (e^+ e^- \to \gamma \gamma)}{dz} =
\frac{2 \pi}{s}\; \left(
\alpha^2  \frac{1+ z^2}{1- z^2}
+ \frac{\alpha}{4}\, s^2\, (1+z^2)  \, \eta
+ \frac{1}{64}\, s^4 \, (1-z^4)\, \eta^2
\right )
\end{equation}
where $z$ ranges from 0 to 1, and $\eta={\cal F}/M_S^4$.
The factor ${\cal F}$ is 
\begin{equation}
{\cal F} = \left\{ 
 \begin{array}{ll}
     \ln \left( \frac{M_S^2}{s} \right ) & {\rm for} \;\; n=2 \;,\\
    \frac{2}{n-2}  & {\rm for} \;\; n>2  \;. 
 \end{array} \right .
\end{equation}
where $n$ is the number of extra dimensions.

The four LEP collaborations have measured the diphoton production
and using the data to constrain the deviation from QED 
by parameterizing the possible deviation from QED 
with a cutoff parameter $\Lambda_\pm$ in the angular distribution:
\begin{equation}
\label{zz}
\frac{d\sigma}{dz} = \frac{2\pi \alpha^2}{s}\; \frac{1+z^2}{1-z^2} \; \left(
1 \pm \frac{s^2}{2 \Lambda_\pm^4}\, (1 - z^2 ) \right )\;.
\end{equation}
The QED cutoff parameter $\Lambda_+$ can be related to $M_S$ by
\begin{equation}
\label{convert}
\frac{M_S^4}{\cal F} = \frac{\Lambda_+^4}{2 \alpha} \;.
\end{equation}
We convert the limit on $\Lambda_+$ to $M_S$ and $M_S$ 
is at most about 1.4 TeV for $n=2$ and about 1 TeV for $n=4$ \cite{eegg}.

The behavior of the new gravity interactions at higher $\sqrt{s}$ can be 
easily deduced from Eq. (\ref{ee}).  The new interaction gives rise to
terms proportional to $s^2/M_S^4$ and $s^4/M_S^8$, which get substantial
enhancement at large $\sqrt{s}$ \cite{eegg}.  The angular
distribution also becomes flatter because in the SM the distribution scales
as 
$(1+z^2)/(1-z^2)$ whereas the terms arising from the new gravity interactions
scale as $(1+z^2)$ and $(1-z^4)$.

\subsubsection{$e^+ e^- \to f\bar f$}

The differential cross section is given by \cite{eeff}
\begin{eqnarray} 
\frac{d \sigma(e^- e^+ \to f \bar f)}{dz} &=& 
\frac{N_f s }{128\pi} \left\{ (1+z)^2 (|M_{LL}(s)|^2 + |M_{RR}(s)|^2 )
+  (1-z)^2( |M_{RL}(s)|^2 + |M_{LR}(s)|^2)  \right .\nonumber \\
&+&
 \pi^2 s^2 ( 1-3z^2 + 4z^4) \eta^2 - 8\pi e^2 Q_e Q_f z^3 \eta 
+ \left.\frac{8\pi e^2}{ s^2_\theta c^2_\theta} \frac{s}{s - M_Z^2} \left(
g_a^e g_a^f \frac{1-3z^2}{2} - g_v^e g_v^f z^3 \right) \eta \right \} 
\nonumber \\
&+& \frac{\delta_{e f} s}{128 \pi}
     \Biggr \{ (1+z)^2 (|M_{LL}(t)|^2 + |M_{RR}(t)|^2 
         + 2 M_{LL}(s) M_{LL}(t) 
+ 2 M_{RR}(s) M_{RR}(t) )  \nonumber \\
&& + 4 ( |M_{LR}(t)|^2 + |M_{RL}(t)|^2 )
+ \frac{\pi^2 s^2}{8} (121 +228z +198z^2 + 84z^3 + 9z^4)\eta^2\nonumber \\
&&-
\frac{\pi s}{2} \eta ( M_{LL}(t) + M_{RR}(t) + M_{LL}(s) + M_{RR}(s) ) (1+z)^2
 (7+z) \nonumber \\
&& + \pi s \eta (M_{LL}(t) + M_{RR}(t) ) (1+z)^2 (1-2z) - 
2 \pi s \eta ( M_{LR}(t) + M_{RL}(t) ) (5+3z)  \Biggr \}
\end{eqnarray}
where
\begin{equation}
M_{\alpha\beta}^{ef}(s) = \frac{e^2 Q_e Q_f}{s} + \frac{e^2}
{s_\theta^2 c^2_\theta}\, \frac{g_\alpha^e g_\beta^f}{s-M_Z^2}  \;,
\end{equation}
$N_f=3(1)$ for quark (lepton) and $z=\cos\theta$, where $\theta$ is the 
scattering angle of the outgoing
fermion.  The latter part of the above equation is for the Bhabha scattering.
The other channels have been studied in Ref. \cite{eeff}.

The results on the sensitivity reach are obtained by combining the likelihood
functions of all these virtual-graviton exchange channels.  The results are
shown in Table \ref{table2}.  The sensitivity is, in fact, dominated by the
Bhabha scattering.

It is clear that the limits obtained by the virtual-graviton
exchange channels are better than those by the real emission channel.  This is
because of the suppression factor $1/M_S^{n+2}$ in the real emission, 
especially severe for large $n$, while no such a suppression factor appears 
in the virtual exchange processes.

Next, 
we compare the sensitivity reach of LC to that at the Tevatron and the LHC.
The diphoton production \cite{greg,eboli2} at hadronic colliders
has been shown very sensitive to the scale $M_S$.  
The sensitivity reach at hadronic machines was estimated using the
diphoton, dilepton channels and their combination \cite{greg}.  At the detector
level, a photon behaves like an electron.  The limits are summarized in the
Table \ref{table3} \cite{greg}.

The sensitivity reach of a mere 0.5 TeV LC with an integrated luminosity
of 10--50 fb$^{-1}$ is far better than the Run II's
of the Tevatron.  An 1 TeV LC with an integrated luminosity of 200 fb$^{-1}$ 
reaches a sensitivity as good as the LHC.  Other studies of large extra
dimensions at $e^+ e^-$ colliders  can be found in Refs. \cite{ee}.

This research was supported in part by the National Center for Theoretical
Science under a grant from the National Science Council of Taiwan R.O.C.

\begin{table}[th]
\caption{\label{table2}\small 
The 95\% C.L. lower limits on the effective Planck scale $M_S$ by combining
all the graviton-exchange processes.}
\medskip
\begin{tabular}{c||cc|cccc}
 &\multicolumn{2}{c|}{$\sqrt{s}=0.5$ TeV} &
  \multicolumn{4}{c}{$\sqrt{s}=1$ TeV} \\
 & ${\cal L}=10$ fb$^{-1}$ &  50 & 10 & 50 & 100 & 200 \\
\hline
$n=3$ & 4.6 & 5.7 & 7.8 & 9.4 & 10.5 & 11.2 \\  
$n=4$ & 3.9 & 4.8 & 6.5 & 
7.9 & 8.9  & 9.4 \\
$n=5$ & 3.5 & 4.3 & 5.9 & 7.2 & 8.0  & 8.5 \\
$n=6$ & 3.3 & 4.0 & 5.5 & 6.7 & 7.5  & 7.9 \\
\end{tabular}

\begin{tabular}{c||cc|cc}
 &\multicolumn{2}{c|}{$\sqrt{s}=1.5$ TeV} &
  \multicolumn{2}{c}{$\sqrt{s}=2$ TeV}\\
 & ${\cal L}=100$ fb$^{-1}$ & 200 &  100  & 200  \\
\hline
$n=3$ & 14.3 & 15.4 &  17.8  & 19.3 \\  
$n=4$ & 12.0 & 13.0 & 15.0  & 16.2 \\
$n=5$ & 10.8 & 11.7 &  13.6  & 14.7 \\
$n=6$ & 10.1 & 10.9 &  12.6  & 13.6 \\
\end{tabular}
\end{table}

\begin{table}[th]
\caption{\label{table3} \small 
Sensitivity to the ADD model parameter $\eta={\cal F}/M_S^4$ 
in Run I, Run II of the Tevatron and at the LHC, using the dilepton, 
diphoton production, and their combination.
 }
\medskip
\centering
\begin{tabular}{c||c|cccccc}
& $\eta_{95}$ (TeV$^{-4}$) & $n=2$ & $n=3$ & $n=4$ & $n=5$ & $n=6$ & $n=7$ \\
\hline
\hline
\underline{Run I (130 pb$^{-1}$)}& & & & & & & \\ 
{}Dilepton & 0.66 & 1.21 & 1.32 & 1.11 & 1.00 & 0.93 & 0.88 \\
{}Diphoton & 0.44 & 1.39 & 1.46 & 1.23 & 1.11 & 1.03 & 0.98 \\ 
{}Combined & 0.37 & 1.48 & 1.53 & 1.29 & 1.16 & 1.08 & 1.02 \\
\hline
\underline{Run IIa (2 fb$^{-1}$)} & & & & & & & \\
{}Dilepton & 0.163 & 1.92 & 1.87 & 1.57 & 1.42 & 1.32 & 1.25 \\
{}Diphoton & 0.077 & 2.40 & 2.26 & 1.90 & 1.71 & 1.60 & 1.51 \\
{}Combined & 0.072 & 2.46 & 2.30 & 1.93 & 1.74 & 1.62 & 1.54 \\
\hline
\underline{Run IIb (20 fb$^{-1}$)} & & & & & & & \\
{}Dilepton & 0.054 & 2.70 & 2.47 & 2.08 & 1.88 & 1.75 & 1.65 \\
{}Diphoton & 0.025 & 3.40 & 3.00 & 2.53 & 2.28 & 2.12 & 2.01 \\
{}Combined & 0.021 & 3.54 & 3.11 & 2.61 & 2.36 & 2.20 & 2.08 \\
\hline
\underline{LHC (14 TeV, 100 fb$^{-1}$)} & & & & & & & \\
{}Dilepton & $2.20\times10^{-4}$ & 10.2 & 9.76 & 8.21 & 7.42 & 6.90 & 6.53 \\
{}Diphoton & $1.24\times10^{-4}$ & 12.1 & 11.3 & 9.47 & 8.56 & 7.97 & 7.53 \\
{}Combined & $1.05\times10^{-4}$ & 12.8 & 11.7 & 9.87 & 8.92 & 8.30 & 7.85 \\
\end{tabular}
\end{table}


\end{document}